Ignoring your neighbors: Moment correlations dominated by indirect or distant interactions in an ordered nanomagnet array

Sheng Zhang,[1] Jie Li,[1] Jason Bartell,[1] Xianglin Ke,[2] Cristiano Nisoli,[3] Paul E. Lammert,[1] Vincent H. Crespi,[1] and Peter Schiffer[1*]

[1]Department of Physics and Materials Research Institute, Pennsylvania State University, University Park, PA 16802, USA

[2]Neutron Scattering Science Division, Oak Ridge National Laboratory, Oak Ridge, TN 37831, USA

[3]Theoretical Division and Center for Nonlinear Studies, Los Alamos National Laboratory, Los Alamos, NM 87545, USA

We have studied the moment correlations within triangular lattice arrays of single-domain co-aligned nanoscale ferromagnetic islands. Independent variation of lattice spacing along and perpendicular to the island axis tunes the magnetostatic interactions between islands through a broad range of relative strengths. For certain lattice parameters, the sign of the correlations between near-neighbor island moments is opposite to that favored by the pair-wise interaction. This finding, supported by analysis of the total correlation in terms of direct and convoluted indirect contributions across multiple pairwise interactions, indicates that indirect interactions and/or those mediated by further neighbors can be tuned to be dominant, with implications for the wide range of systems composed of interacting nanomagnets.

PACS numbers: 75.50.Lk, 75.30.-m, 75.75.-c

*Corresponding author: Peter Schiffer, pes12@psu.edu

The study of nanoscale single-domain ferromagnets has revealed a range of fascinating behavior, both in the collective properties of superparamagnetic nanoparticles produced through chemical synthesis [1] and in the local properties of lithographically fabricated single-domain ferromagnetic islands or wires arranged so that the magnetostatic interactions are frustrated [2]. Strong shape anisotropy makes the magnetic moments in these lithographically defined systems analogous to Ising spins. Unlike atomic-scale Ising spins, however, the islands' moments can be measured both locally and collectively and tuned by changing the lattice geometry. Several different frustrated lattices have been studied, including square [2,3,4], triangular [5], hexagonal (i.e. kagome) [6,7,8,9,10,11,12,13] and brickwork lattices [6], as well as isolated clusters of islands [14].

In this Letter, we study triangular lattice arrays of islands [5], with all of the islands' long axes and thus the magnetic moments co-aligned, as shown in Fig. 1. The ability to separately control the inter-island spacing along the two axes perpendicular and parallel to the islands' common axial direction, combined with the (dipole-like) anisotropy of the island-island interaction, provides an opportunity to separately vary the relative magnitudes and even the signs of the interactions between different classes of Ising-like moment pairs without altering the lattice geometry. Through such variation, which is not possible in higher-symmetry geometries (e.g. square, hexagonal), we explicitly demonstrate that indirect and further-neighbor interactions can control the correlations between neighboring moments. The results, which can be described within a self-consistent Ornstein-Zernike equation, elucidate the complexities of interactions among nanomagnet assemblies, with relevance to the wide range of such assemblies currently under study.

Fig. 1(a) shows a schematic of a triangular lattice of ferromagnetic islands fabricated using electron-beam lithography and MBE-deposited permalloy ($Ni_{0.81}Fe_{0.19}$). The islands are lithographically defined to be 220 nm long, 80 nm wide and 25 nm thick; the elongated shape ensures that the islands are single-domain with a magnetic moment along the long (*i.e.*, easy) axis due to shape anisotropy. We call the inter-island spacing within a row of islands *x*, and the spacing between rows *d*, as shown in Fig. 1(a). We can tune the lattice geometry by changing both *d* and *x*, thus controlling the relative strengths of different interactions. We examined several series of arrays with varying lattice parameters. Four of these series had a fixed ratio of *d/x* (0.4, 0.8, 1.0 and 1.2) with each series including *x* = 400, 480, 560, 680, and 880 nm. We also studied a series with *x* = 400 nm and varying *d*, and a series with *d* = 200 nm and varying *x*.

Samples were prepared for study through a process of ac demagnetization that enables the system to access low-energy states, as described in detail elsewhere [15,16]. The sample is rotated at 1000 rpm in a magnetic field starting well above the single-island coercive field and stepped down in 1.6 Oe increments, reversing polarity at each step and holding steps for 5 seconds. After demagnetization, the island moments are imaged via magnetic force microscopy (MFM) in patches at five locations within each array. The typical MFM image shown in Fig. 1(b) demonstrates the single-domain nature of the islands, in that each island has two halves with black and white representing the north and south poles respectively. The number of islands imaged at each location varies from 300 to 1200 depending on the lattice spacing, and the error bars in the figure are the standard errors obtained after averaging data from the five different MFM images taken at different locations within each pattern. The data were consistent with those published previously on a more limited range of triangular lattices [5], showing small domains of moment ordering for certain lattice spacings.

We examined correlations between the moment orientations for distinct types of nearest-neighbor pairs denoted 1, 2, 3 and 4 in Fig. 1(a). We assign island $n$ a spin $S_n$, equal to +1 or -1, corresponding to its moment pointing up or down. The correlation $C_t$ for pair type $t = 1, 2, 3$ is then the experimental average of $S_n \cdot S_m$ over all pairs $(n,m)$ of the given type [17]. The magnetic interaction energy between a type-t pair $(n,m)$ is written $-J_t S_n \cdot S_m$. We calculated the interaction energies $J_t$ using the OOMMF micromagnetic simulation package [18]. A simple point-dipole approximation suffices to understand qualitative trends with lattice parameters, in particular the change in sign of $J_2$, but such an approximation is not very accurate, especially at small lattice spacings where the extended nature of the islands becomes important. If the correlation $C_t$ is predominantly controlled by the interaction $J_t$, then we would expect $C_t$ and $J_t$ to have the same sign. Naively, one also expects the magnitudes of $C_t$ and $J_t$ to vary with the lattice parameters in qualitatively similar ways. We therefore plot $Sgn(J_t)C_t$ in the figures below, where positive values of the product indicate that $C_t$ and $J_t$ have the same sign.

Fig. 2 plots $Sgn(J_t)C_t$ and the interaction energies $J_t$ for the three nearest neighbor types ($t = 1, 2, 3$) as a function of $x$ for the samples with fixed ratios of $d/x$, with separate plots for each ratio. The case of $d/x = 0.4$ is distinct from the other ratios, in that $J_2$ changes sign from negative to positive with increasing $x$ and the magnitude of $J_3$ is much larger than all other interactions. For the other three lattice ratios, $J_3$ has much smaller magnitude than $J_1$ and $J_2$ and none of the interaction energies change sign. Furthermore, while $J_1$ and $J_2$ show the same qualitative decrease in magnitude with increasing $x$, $|J_1|$ is larger than $|J_2|$ for $d/x = 0.4$ and 1.2, while it is smaller for $d/x = 0.8$ and 1.0. This variability of the relative magnitudes of different interactions is quite different from the behaviors of the square, hexagonal and brickwork geometries previously studied [6]: in those cases, the strongest interaction is always between one fixed type of pair.

Figs 3(a,b,c) shows the correlations $Sgn(J_t)C_t$ for $t = 1, 2, 3$ respectively as a function of the row spacing $d$ for all samples. The calculated $J_1$, $J_2$ and $J_3$ are shown for comparison in Figs 3(d,e,f). For the type-1 neighbors shown in Figs 3(a,d) the interaction $J_1$ is always negative and depends only on $x$. $Sgn(J_1)C_1$ is positive for almost all samples, as expected if the direct interaction controls the correlation. However, several points have negative $Sgn(J_1)C_1$, indicating that the correlations are *opposite* what would be expected from the direct interactions between the moments.

The type-2 neighbors are unusual in that $J_2$ may take either sign depending on $x$ and $d$. Figs 3(b,e) show that $C_2$ mostly tracks the $d$ dependence of $J_2$: weak at small $d$, reaching a positive maximum at intermediate $d$ and then approaching zero at large $d$. However, the negative $Sgn(J_2)C_2$ in Fig. 3(b) (and also Fig. 2(a)) provides another case where the observed correlations are *opposite* to what would be expected from the direct interactions between the moments. Fig. 3(c) shows a monotonic decrease of $Sgn(J_3)C_3$ with $d$, with no indication of $x$ dependence, mirroring the behavior of $J_3$ and positive for all samples.

We now consider in more detail the anomalous samples for which $Sgn(J_t)C_t$ is negative, and in which the direct interactions between moments are dominated by other interactions, either indirect or further-neighbor. Consider first the negative values of $Sgn(J_1)C_1$. Fig. 1(a) shows how a triangle of three nearby islands can be constructed with one $J_1$ edge and two $J_2$ edges. Any indirect interaction which traverses a chain of two equivalent spin pairs – like the two $J_2$ edges – is effectively ferromagnetic. Hence in the limit of $J_2 \gg J_1$ we expect $C_1$ to be positive (and $Sgn(J_1)C_1$ negative) due to a dominant indirect ferromagnetic interaction mediated by $J_2$. Indeed, this is what we observe in Fig. 2(b) and Fig. 3(a) for small $x$. The large $J_3$ has little effect, since the vertical ferromagnetic chains favored by $J_3$ are entirely compatible with this spin texture.

The negative values of $Sgn(J_2)C_2$ in Figs 2(a) and 3(b) cannot be explained by the influence of the near-neighbor interactions $J_1$ or $J_3$ since the subsets of islands mutually coupled by $J_1$ and $J_3$ form two non-overlapping sublattices that do not contain $J_2$. On the other hand, they might be attributable to the influence of further neighbor interactions – in particular the $J_4$ interaction shown in Fig. 1(a). As shown in Fig. 2(a), at $d/x=0.4$ where $Sgn(J_2)C_2$ becomes negative, both $J_3$ and $J_4$ favor ferromagnetic alignment of the respective moment pairs whereas $J_2$ favors antiferromagnetic alignment. Therefore a triad with one $J_2$, one $J_3$ and one $J_4$ interaction (shown in Fig. 1(a) as red, blue and dashed green lines) is frustrated and will accept a negative $Sgn(J_2)C_2$ so long as $J_3$ and $J_4$ dominate over $J_2$. For $d/x = 0.4$, the vertical interaction $J_3$ is by far the strongest, while $J_4$ is either comparable to or larger than $J_2$ precisely when $C_2$ and $J_2$ are opposite in sign.

These qualitative arguments can be placed on a more quantitative footing by decomposing the total inter-island correlations into direct and indirect contributions. This will lead to a rough criterion, given the important pair interaction energies, for when to expect a specific pair correlation to disagree in sign with the corresponding interaction. The states of the island moments are certainly not in thermal equilibrium – interaction energies between neighboring islands are $\sim 10^4$ K. Furthermore, given the highly dissipative process through which they achieve their final states, barriers to flipping the moments may be as important as the interaction energies themselves. On the other hand, a rough proportionality between shifts of well depths and barriers is common in models of crossing barriers between energy wells, such as the Butler-Volmer theory of electrode processes [19]. In addition, for square-lattice artificial spin ice, a model with a single effective temperature fits the behavior fairly well [20]. Nevertheless, regardless of the thermodynamic state of the lattice, we are interested in how the interactions of two islands each with a third influences the correlation between the first two. The Ornstein-Zernike equation, which figures prominently in the theory of liquid structure, concerns precisely that problem [21]. It expresses the total correlation function as the sum of the direct correlation function and the convolution of the direct correlation with the total correlation. Such a relation is of a general probabilistic character and is not limited to genuine thermal equilibrium.

We consider as an *ansatz* that the direct correlation function is proportional to the interaction energies $J_t$ for pairs of type-$t$, $t = 1, 2, 3, 4$, and otherwise zero. This corresponds to the mean spherical approximation in liquid state theory. For $t = 1$ and 2, the spin Ornstein-Zernike equation [22] then gives:

$C_1 \propto J_1 + 2C_2J_2 + 2C_4J_4$

and

$C_2 \propto J_2 + C_2J_1 + C_1J_2 + C_2J_3 + C_3J_2 + C_3J_4 + C_4J_3,$

where the additional terms arise from indirect interactions and interactions with further neighbors. For a state in thermal equilibrium, the proportionality constant in these expressions would be the inverse temperature. Fig. 4 plots $C_1$ and $C_2$ against our *ansatz* sum of interaction energies. Neither plot is a simple straight line that would indicate a single effective temperature (assuming uniform applicability of the mean-spherical approximation). However, somewhat different effective temperatures for different lattice parameters and even for different interactions would not be unexpected. The striking feature of these plots is the way the data funnel through the origin, *i.e.* the *ansatz* sum of interaction energies appears to control the sign of the correlations. Although a somewhat crude test, this result does provide strong corroboration of the hypothesis that indirect interactions explain the disagreements in sign between $C_1$ and $J_1$ and between $C_2$ and $J_2$.

While interactions between nanomagnets are important to understanding both lithographically fabricated systems and assemblies of ferromagnetic nanoparticles [1, 23], previous analyses have focused on the direct interactions between the moments. The general phenomenon of frustration implies that interactions will compete, suggesting that correlations between moments can be opposite from the sign of their direct interactions, and such phenomena are routinely expected in spin glasses [24] and even can be seen in previous data on nanomagnet arrays. In the higher symmetry artificial frustrated systems [2, 6] examined to date, however, the relative strengths of the various interactions are fixed by the specific lattice spacing and geometry, and thus the nature of such opposite correlations cannot be tuned. In systems of frustrated magnetic materials (i.e., systems in which interactions between atomic moments are frustrated), tuning is even more difficult, as is the direct probing of individual local interactions. Our triangular lattice of nanomagnets offers special opportunities in that we can vary the ratios of the different interactions while also changing the strengths of the interactions, and we have four different neighbor interactions that are significant. The lower-symmetry triangular lattice thus allows for a detailed examination of the indirect and further-neighbor interactions and is an explicit demonstration of cases in which these interactions are dominant. The Ornstein-Zernike framework that we develop then elucidates how the different interactions contribute to produce the experimentally observed correlations. The ability to "ignore one's neighbors" is a self-consistent, collective property, as embodied in the spin Ornstein-Zernike framework. Characteristics of the interaction topology that favor this phenomenon include multiple symmetry-distinct pairwise hopping paths between a given island pair where the multiplicity of longer, indirect paths counter-balances the stronger net interactions along shorter paths.

Further-neighbor and indirect interactions among nanomagnets are likely to be of increasing interest in future studies of more complex and less symmetric structures. In particular, the design of structures that allow the application of local magnetic fields to influence moment configurations will need to carefully consider these sorts of interactions. They may also be of considerable interest in possible device applications that rely upon magnetostatic interactions between components [ 25 ], and in detailed modeling of three-dimensional nanomagnet assemblies [1,23].

This research was supported by the U.S. Department of Energy, Office of Basic Energy Sciences, Materials Sciences and Engineering Division under Award # DE-SC0005313, and by the Army Research Office and the National Science Foundation MRSEC program (DMR-0820404), the National Nanotechnology Infrastructure Network and an REU Supplement to NSF grant DMR-0701582. We greatly appreciate Professor Chris Leighton and Mike Erickson for sample preparation and helpful discussions.

Figure 1. (a) Schematic of triangular lattice arrays. The near-neighbor pair types are labeled $C_1$, $C_2$, $C_3$, and $C_4$. (b) MFM image of a section of one triangular lattice array ($x = 680$ nm, $d/x = 0.4$). Black and white halves represent the north and south poles of each ferromagnetic island. Scale bar is 1 μm.

Figure 2. (a-d) Correlation values $Sgn(J_t)C_t$ as a function of $x$ for fixed $d/x$ ratios 0.4, 0.8, 1 and 1.2 respectively. (e-h) Interaction energies $J_t$ as a function of $x$ for fixed $d/x$ ratios 0.4, 0.8, 1 and 1.2 respectively, as simulated by OOMMF software (in units of $10^{-19}$ Joules).

Figure 3. (a-c) Correlation values $Sgn(J_1)C_1$, $Sgn(J_2)C_2$, $Sgn(J_3)C_3$ as a function of $d$ for fixed $x$ distance. Each contains 5 groups of patterns with fixed $x$ (400 nm, 480 nm, 560 nm, 680 nm, 880 nm). (d-f) Interaction energies $J_1$, $J_2$, $J_3$ as a function of $d$ for fixed $x$, as simulated by OOMMF software (in units of $10^{-19}$ Joules).

Figure 4. Tests of the *ansatz* (in units of $10^{-19}$ Joules) that further neighbor and indirect interactions are important in determining $C_1$ and $C_2$. The data funnel through the origin in both plots. The fan-out away from the origin may be attributable to different effective temperatures in different lattices.

Figure 1

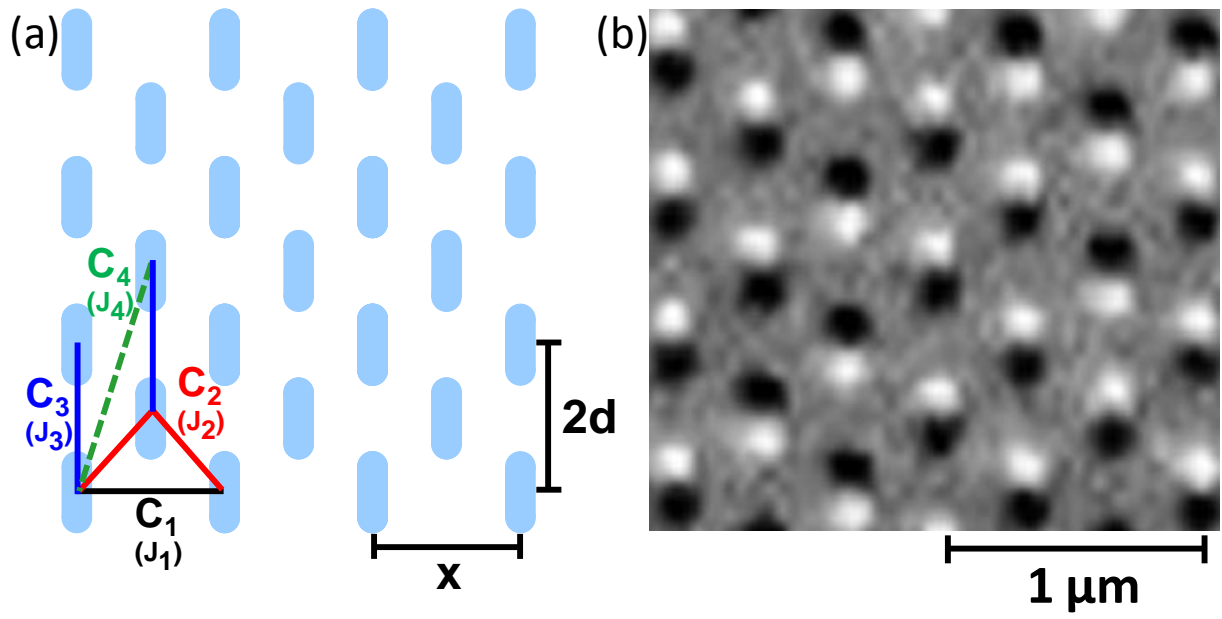

Figure 2

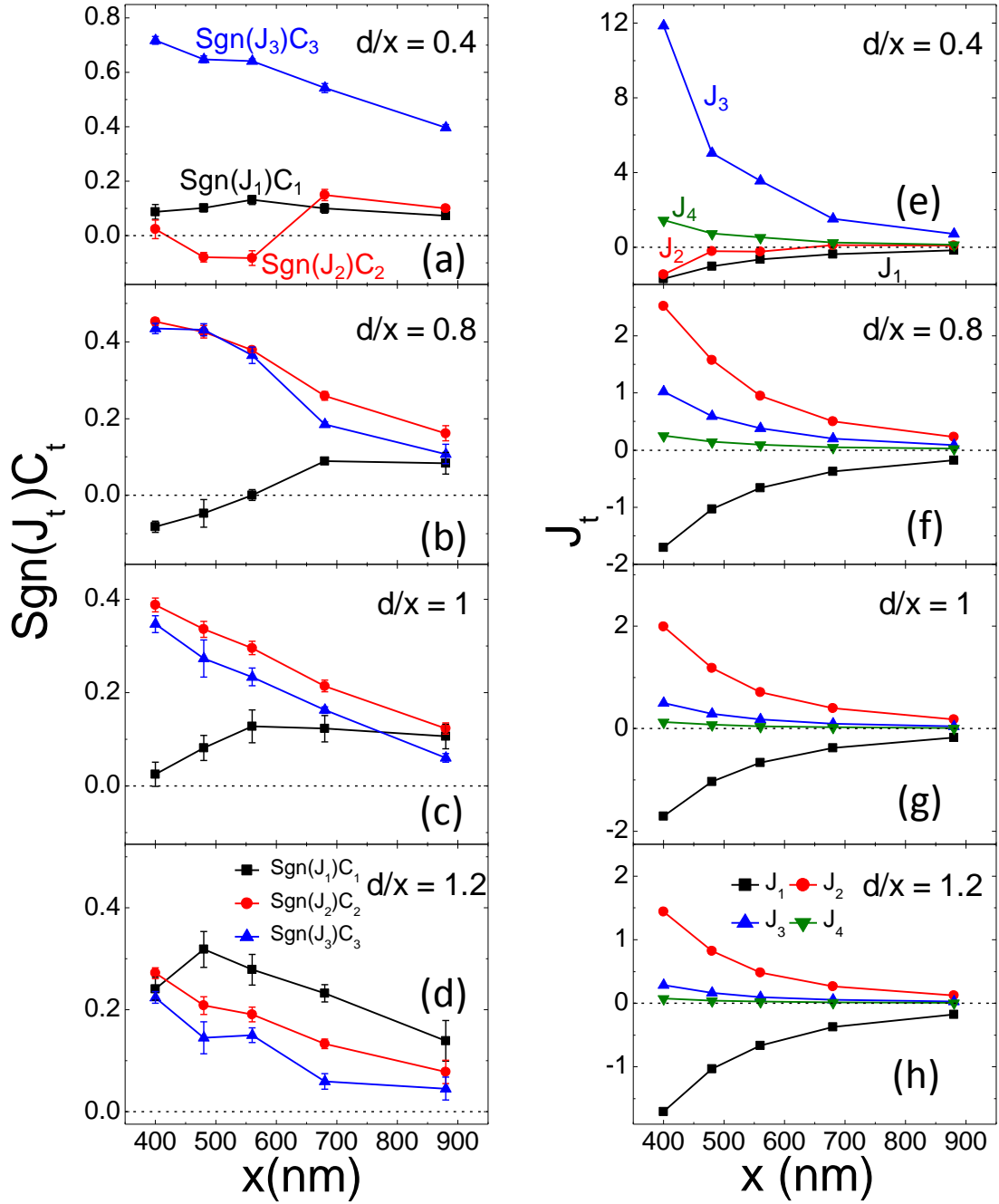

Figure 3

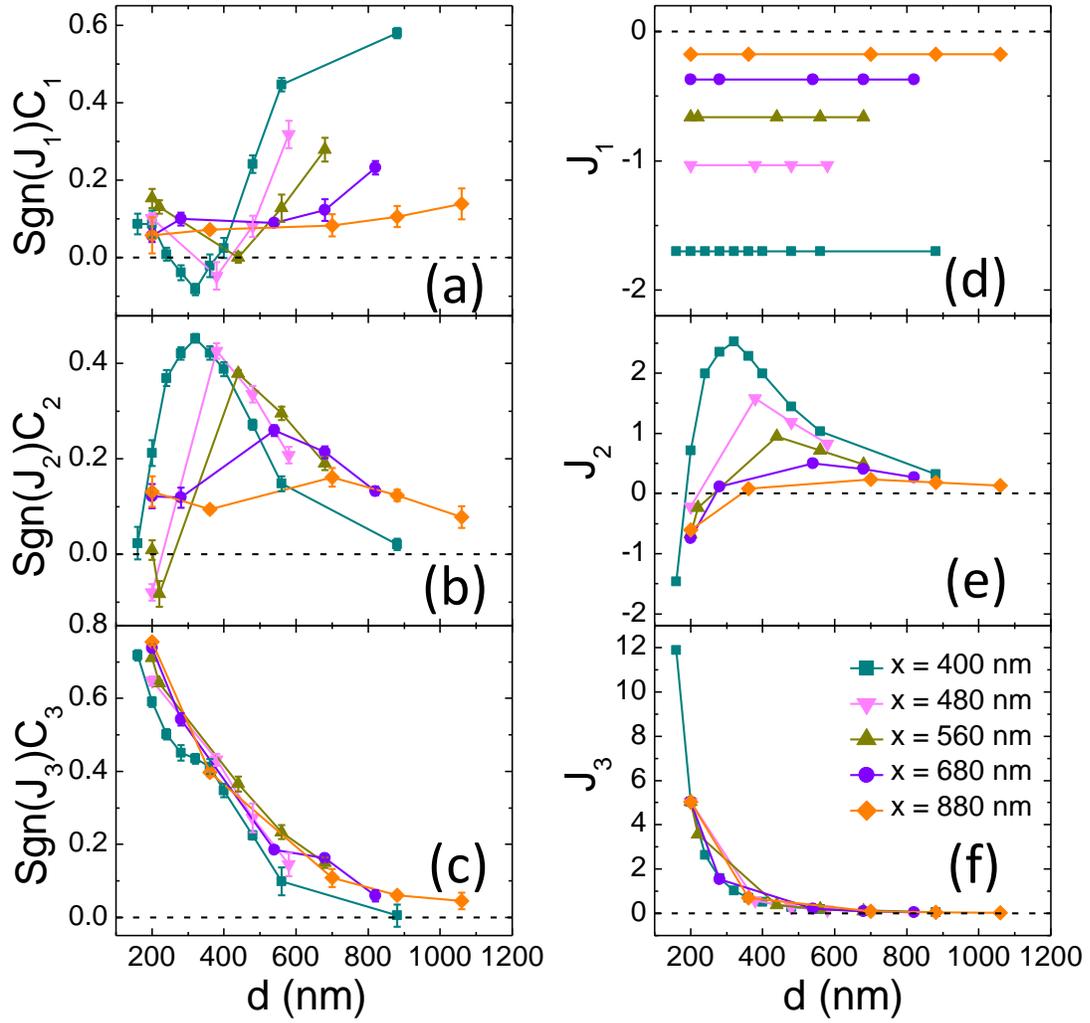

Figure 4

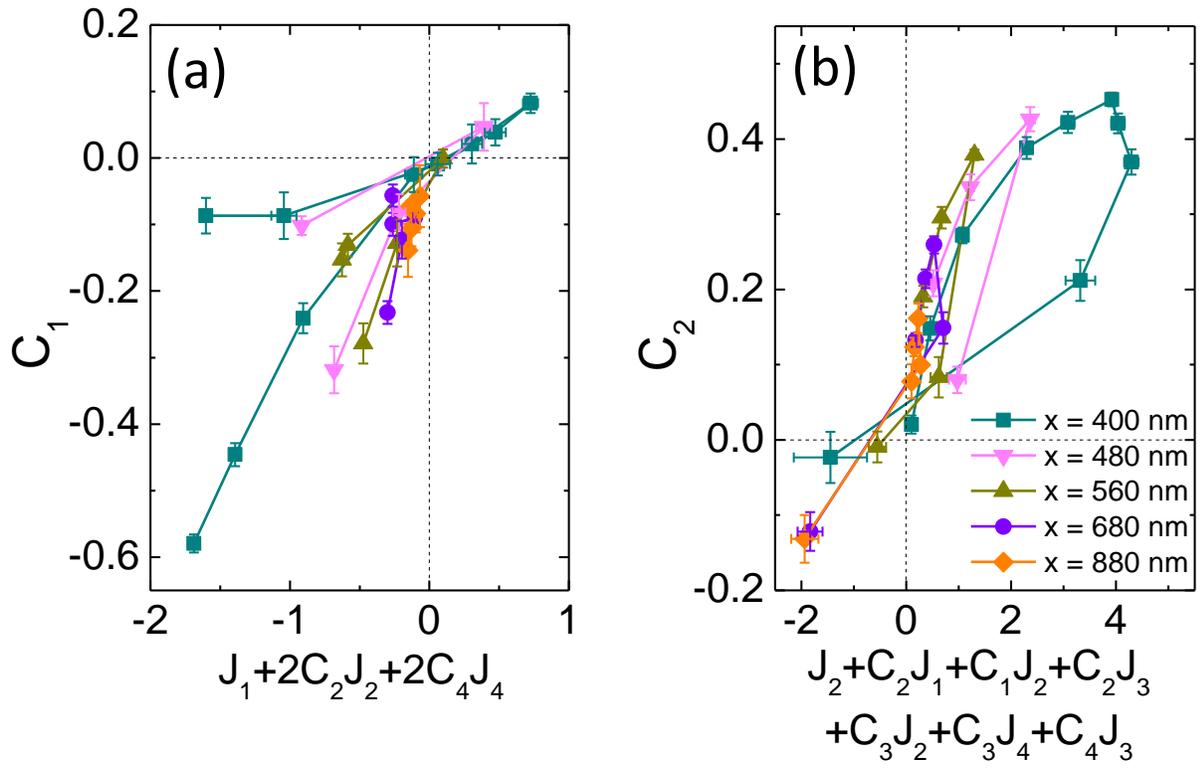